\documentclass[aps,prd,twocolumn,groupedaddress,nofootinbib,superscriptaddress]{revtex4-1}

\usepackage{natbib}
\usepackage{graphicx}
\usepackage{color}
\usepackage{amsmath, amsfonts, amssymb}
\usepackage{hyperref}
\usepackage{subfigure} 
\usepackage{amsmath}
\usepackage{ulem}

\begin{document}

 \title{Prospects for intermediate mass black hole binary searches with advanced gravitational-wave detectors}

 \author{G. Mazzolo}
 \affiliation{Albert-Einstein-Institut, Max-Planck-Institut f\"{u}r Gravitationsphysik, D-30167 Hannover, Germany}
 \affiliation{Leibniz Universit\"{a}t Hannover, D-30167 Hannover, Germany}
 \author{F. Salemi}
 \affiliation{Albert-Einstein-Institut, Max-Planck-Institut f\"{u}r Gravitationsphysik, D-30167 Hannover, Germany}
 \affiliation{Leibniz Universit\"{a}t Hannover, D-30167 Hannover, Germany}
 \author{M. Drago}
 \affiliation{Universit\`a di Trento, I-38123 Trento, Italy}
 \affiliation{INFN, TIFPA, Trento, Italy}
 \author{V. Necula}
 \affiliation{University of Florida, Gainsesville, FL 32611, USA}
 \author{C. Pankow}
 \affiliation{University of Wisconsin Milwaukee, Milwaukee, WI 53201, USA}
 \author{G. A. Prodi}
 \affiliation{Universit\`a di Trento, I-38123 Trento, Italy}
 \affiliation{INFN, TIFPA, Trento, Italy}
 \author{V. Re}
 \affiliation{INFN, Sezione di Roma Tor Vergata and Universit\`a di Roma Tor Vergata, I-00133 Roma, Italy}
 \author{V. Tiwari}
 \affiliation{University of Florida, Gainsesville, FL 32611, USA}
 \author{G. Vedovato}
 \affiliation{INFN, Sezione di Padova and Universit\`a di Padova, I-35131 Padova, Italy}
 \author{I. Yakushin}
 \affiliation{LIGO - Livingston Observatory, Livingston, LA 70754, USA}
 \author{S. Klimenko}
 \affiliation{University of Florida, Gainsesville, FL 32611, USA}

 \begin{abstract}
We estimated the sensitivity of the upcoming advanced, ground-based gravitational-wave observatories (the upgraded LIGO and Virgo and the KAGRA interferometers) to coalescing intermediate mass black hole binaries (IMBHB). We added waveforms modeling the gravitational radiation emitted by IMBHBs to detectors' simulated data and searched for the injected signals with the coherent WaveBurst algorithm. The tested binary's parameter space covers non-spinning IMBHBs with source-frame total masses between 50 and 1050 $\text{M}_{\odot}$ and mass ratios between $1/6$ and 1$\,$. We found that advanced detectors could be sensitive to these systems up to a range of a few Gpc. A theoretical model was adopted to estimate the expected observation rates, yielding up to a few tens of events per year. Thus, our results indicate that advanced detectors will have a reasonable chance to collect the first direct evidence for intermediate mass black holes and open a new, intriguing channel for probing the Universe over cosmological scales.
\end{abstract}
 \maketitle

 \section{Introduction} \label{Introduction}
Intermediate mass black holes are an elusive class of black holes with masses between few tens and $\sim 10^5$ solar masses \cite{Coleman_Miller2}. These objects might be of relevance for explaining supermassive black hole formation \cite{Sesana, Volonteri1}, star-cluster dynamics \cite{Ferraro, Gebhardt, Noyola, Trenti, van_den_Bosch, Bash} and ultra-luminous X-ray sources \cite{Farrell, Kaaret, Kajava, Strohmayer1, Strohmayer2, Vierdayanti}. However, to date photon-based astronomy has delivered only ambiguous indications for their existence \cite{Coleman_Miller2}. Direct evidence could be provided by gravitational-wave (GW) astronomy, as coalescing intermediate mass black hole binaries (IMBHBs) are expected to be the brightest sources of gravitational radiation accessible to ground-based GW observatories.

Several black-hole binary searches have been undertaken on data collected by the most sensitive GW detectors operating in the past years, the LIGO and Virgo interferometers \cite{Abbott1, Accadia2}. Most of the searches have been performed with analysis methods based on matched-filtering \cite{Abbott2, Abbott3, Abbott4, Abadie1, Abadie2, Abadie3, Aasi1,Aasi4}. This approach requires the generation of potentially large filter banks covering the investigated parameter space  \cite{Sathyaprakash1,Allen,Babak}. However, the most generic gravitational waveform from compact binaries depends on several parameters \cite{Maggiore} and a bank of filters accurately modeling all possible signals is currently lacking. Thus, searches have also been conducted with unmodeled approaches \cite{Abadie4,Aasi3}, which are sensitive only to excesses of signal power and do not require accurate knowledge of the waveform \cite{Anderson}. Due to the lack of signal constraints, unmodeled searches are typically more susceptible than matched-filtering to triggering by environmental and instrumental glitches. However, the two approaches show comparable sensitivity when the in-band portion of the signal spans a limited area in the time-frequency domain \cite{Anderson}, as in the case of IMBHB searches performed with ground-based interferometric detectors \cite{Aasi5}.

Thus far, no black-hole binary has been discovered in LIGO-Virgo data. In particular, the upper limits placed on the IMBHB merger-rate density are a few orders of magnitude above rough predictions based on astrophysical models \cite{Abadie5}. However, the ability to detect GWs from compact binaries is expected to increase significantly in a few years, when the advanced LIGO, Virgo and KAGRA detectors will come online \cite{Harry, Accadia, Somiya, Aasi2}. 

In this paper we assess the sensitivity of unmodeled IMBHB searches with advanced detectors. Search ranges and expected observation rates are estimated for different detector networks over a wide IMBHB parameter space. The plan of the paper is as follows: Section \ref{analysis_overview} presents an overview of the analysis; the results are reported in Section \ref{search_results}$\,$; the main sources of uncertainty impacting the analysis are discussed in Section \ref{discussion}$\,$; the conclusions are outlined in Section \ref{conclusion}$\,$.
 \section{Analysis overview} \label{analysis_overview} 

\subsection{Advanced detectors} \label{Advanced_detectors}
The next few years will see the initial operations of the advanced LIGO, Virgo and KAGRA detectors \cite{Aasi2}. Advanced LIGO will consist of two interferometers with 4-km arms located at the same sites as the previous LIGO detectors, i.e., in Hanford, Washington (hereafter denoted H) and Livingston, Louisiana (L)\footnote{The installation of a third LIGO detector in India (IndIGO) is in the final stages of consideration by the Indian funding agency \cite{Indigo, Iyer}. Assuming that the final funding approvals will be granted, the first runs are expected for 2020. The design sensitivity, equivalent to that of H and L, should be reached by no earlier than 2022 \cite{Aasi2}. As detector site and orientation had not been chosen at the time of writing, the IndIGO interferometer was not considered for this analysis.},  and will soon start collecting data (2015) \cite{Harry}. Advanced Virgo (V, 3-km arms) will also operate in the same location as the previous instrument (Pisa, Italy) and, depending on the sensitivity attained, might also start operating in 2015 \cite{Accadia}. The site chosen for the KAGRA detector (K, 3-km arms) is Hida, Japan, and the first science runs are scheduled for 2018 \cite{Somiya}.

\begin{figure}[t!]
 \begin{center}
 \includegraphics[width = 9.0 cm, height = 7.0 cm]{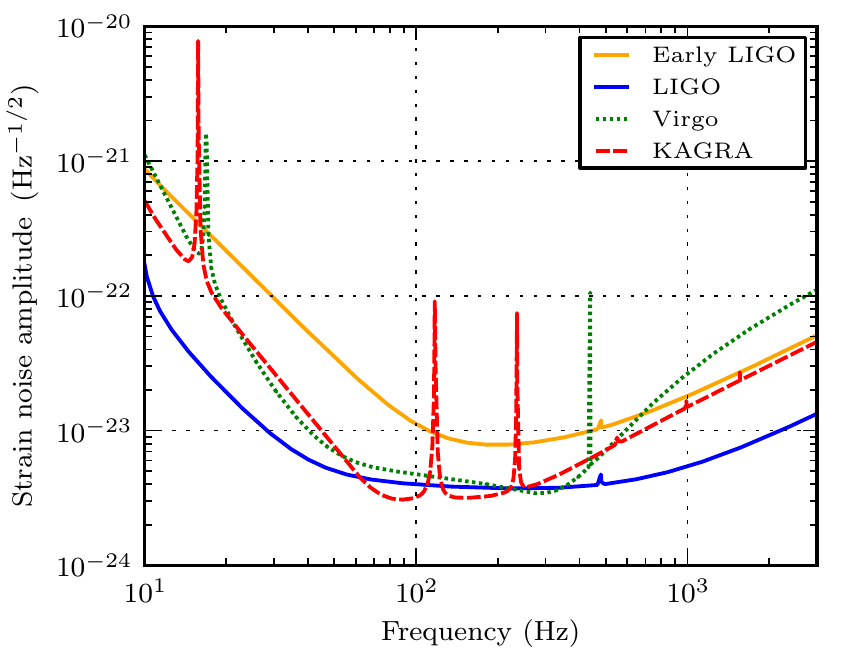}
 \caption{(color online) Design strain sensitivity for advanced detectors between 10 Hz and a few kHz. The sensitive band for intermediate mass black hole binaries resides below $\sim 100$ Hz.} 
 \label{sensitivity} 
 \end{center}
\end{figure}

The detectors' design sensitivities, reported in Figure \ref{sensitivity}$\,$, are based on different models of the noise sources, e.g. in the lower frequency band \cite{Harry, Accadia, Somiya}, and will be progressively achieved \cite{Aasi2}. Figure \ref{sensitivity} also shows the sensitivity at which the first advanced LIGO science runs should be performed (Early LIGO). Relative to the previous LIGO-Virgo observatories, the advanced detectors target a significant sensitivity improvement over the whole bandwidth (about one order of magnitude in the most sensitive band, around $\sim 200$ Hz), together with an extension of the viable low-frequency band. The improvement at low frequencies will be particularly relevant for IMBHB searches, as it will extend the upper end of the binary total-mass spectrum beyond that accessible with past GW interferometers \cite{Abbott4,Aasi4,Abadie4,Aasi3}. 

\subsection{Data-analysis algorithm} \label{data_analysis_algorithm}
This analysis was conducted with coherent WaveBurst, the data-analysis algorithm used for the past LIGO-Virgo unmodeled black-hole binary searches \cite{Abadie4,Aasi3}. Coherent WaveBurst performs a coherent analysis on data from multiple detectors \cite{Klimenko1}. After decomposing the data into a time-frequency representation, the algorithm identifies coherent triggers from regions in the time-frequency domain with excess power relative to the noise level \cite{Klimenko4}. Network events are subsequently reconstructed in the framework of a constrained maximum-likelihood analysis \cite{Klimenko2,Klimenko3}. For this analysis, we applied a further, weak constraint to favor the reconstruction of elliptically-polarized waveforms \cite{Pankow}. For each event, the algorithm reconstructs the GW amplitude and the source sky position, together with the calculation of a number of coherent statistics, see Appendix \ref{appendix_A}$\,$.

\subsection{Simulation procedure}
We estimated the sensitivity to IMBHBs of three detector networks: HKLV, HLV and HL$\,$. The HL network was tested with the Early LIGO configuration as well. The Early HL network is expected to be representative of the initial science runs, the other networks of the runs performed at the design sensitivity. 

The sensitivity of the tested networks was estimated via Monte Carlo detection-efficiency studies. Waveforms modeling GWs from IMBHBs were added (``injected'') into simulated Gaussian stationary noise colored to resemble the design sensitivities in Figure \ref{sensitivity}$\,$, and searched for with coherent WaveBurst. We used the non-spinning EOBNR waveform family \cite{Buonanno1, Pan}. This includes the leading $(l,m) = (2,\,2)$ mode and the higher order, sub-dominant $(2,\,1)$, $(3,\,3)$, $(4,\,4)$ and $(5,\,5)$ modes. 

The simulated waveforms were distributed uniformly in binary source-frame total mass and mass ratio from 50 to 1050 $\mbox{M}_{\odot}$ and 1/6 to 1$\,$, respectively. The uniform distributions were motivated by the lack of astrophysical constraints on the actual distributions of the IMBHB parameters. The considered total-mass and mass-ratio ranges include the IMBHBs to which the advanced detectors show the greatest sensitivity, see Section \ref{search_results}$\,$. The sensitivity to less massive systems decreases due to the progressively broader time-frequency area spanned by the inspiral stage within the detectors' bandwidth. More massive systems become rapidly inaccessible due to the steep low-frequency increase of the design strain noise amplitude (Figure \ref{sensitivity}$\,$). The tested mass-ratio interval was selected based on the range over which the EOBNR waveforms had been calibrated to numerical-relativity simulations \cite{Pan}. The waveforms were uniformly distributed in the binary's inclination with respect to the line of sight and in comoving volume. Note that the adopted total-mass and spatial distributions refer to source-frame and comoving values, respectively. As advanced detectors are expected to be sensitive to IMBHBs up to a few Gpc, see Section \ref{search_results}$\,$, redshift effects on observed binary parameters such as masses and distances must be taken into account, see Appendix \ref{appendix_B}$\,$.

The simulation studies enabled the calculation of the comoving visible volume $V_{\text{vis}}$ for IMBHB mergers. Following the procedure in \cite{Abadie4, Aasi3}, the $V_{\text{vis}}$ was calculated as
\begin{equation} \label{visible_volume}
 V_{\text{vis}} \left(m_1, m_2 \right) = \sum_{i} \frac{1}{\rho_i} = \sum_{i} \frac{4 \pi r_i^2}{\frac{dN_{\text{inj}}}{dr} \left(r_i \right)} \ .
\end{equation}
Here $m_1$ and $m_2$ are the source-frame masses of the two binary components, the sum runs over the recovered injections, $\rho_i$ is the density number associated to each injection, $r_i$ the comoving distance to the sources and $dN_{\text{inj}}/dr$ their radial distribution. We assessed the sensitivity to IMBHBs of advanced detectors in terms of the effective radius $R_{\text{eff}}\,$, defined as the radius of the sphere with volume $V_{\text{vis}}\,$.

The HKLV recovered injections were selected by applying a threshold of 11 on the reconstructed network signal-to-noise ratio (SNR). This was chosen somewhat arbitrarily to apply a threshold comparable to the lowest reconstructed network SNR at which simulated signals were recovered by a previous LIGO-Virgo, four-detector IMBHB search \cite{Abadie4}. The threshold selection was based on the fourfold configuration as this will be the most sensitive network for searches conducted on real detectors data, see Section \ref{search_ranges}$\,$. For the sake of simplicity, the same threshold was applied to the other tested networks. The threshold selection procedure followed in this paper differs from the approach considered for the previous coherent WaveBurst IMBHB searches, see Appendix \ref{appendix_A} and \cite{Abadie4, Aasi3}. It was adopted as no realistic estimate of the background affecting searches conducted with advanced detectors was available at the time of the analysis. Nevertheless, we show in Section \ref{search_results} how the $R_{\text{eff}}$ varies over a wide range of different thresholds.

The results presented in the following sections are expressed in terms of comoving distance and as a function of the source-frame companion masses, and averaged over the sky positions and inclinations of the binary systems.
 \section{Results} \label{search_results}

\subsection{Ranges} \label{search_ranges}

\begin{figure*}[htp]
 \centering
 \subfigure[]{\includegraphics{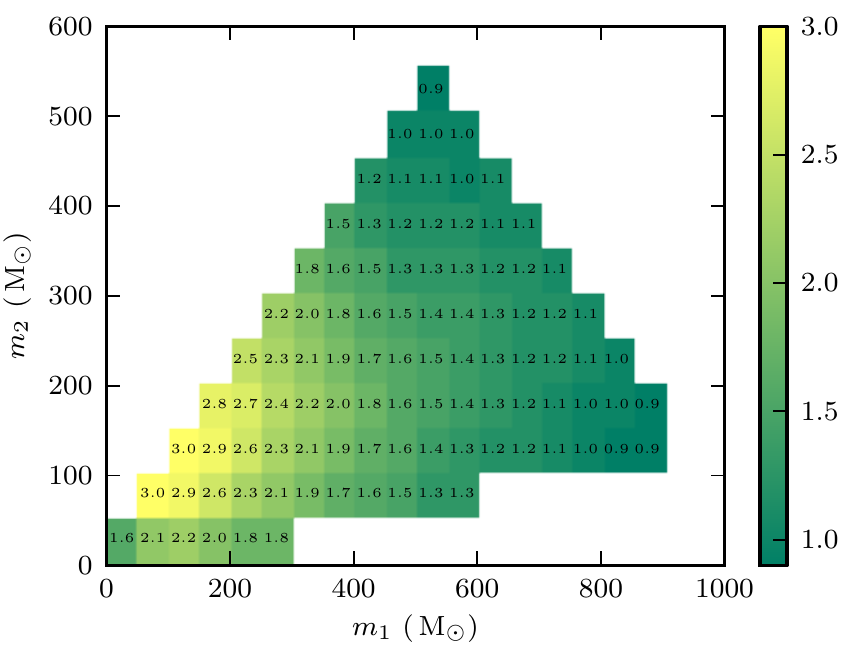}}
 \hspace{3mm} 
 \subfigure[]{\includegraphics{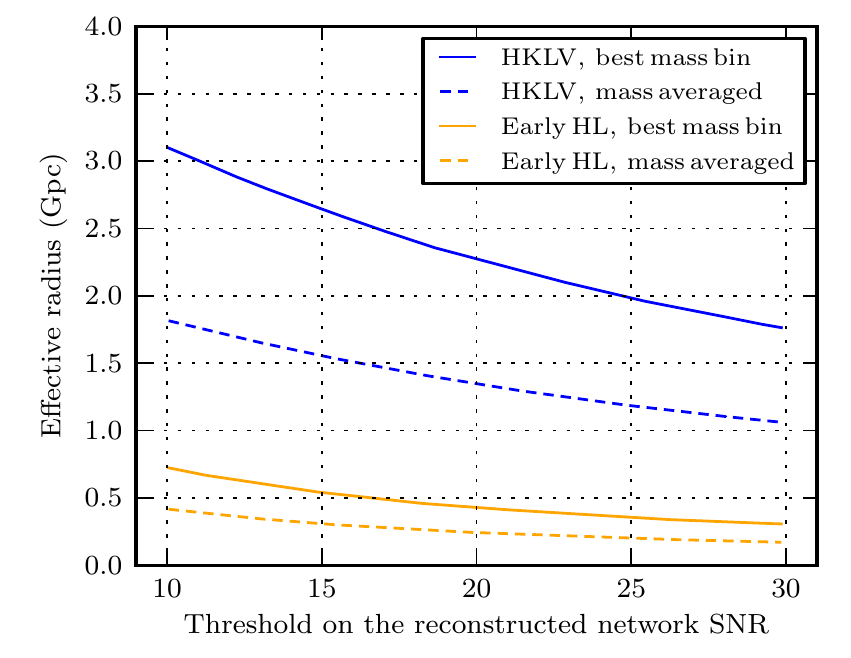}}
   \caption{(color online) a) Effective radius in Gpc calculated for the HKLV network as a function of the source-frame companion masses. b) Dependence of the HKLV and Early HL effective radii on the reconstructed network SNR. The solid lines are calculated for the most sensitive mass bin, centered at $125 + 125 \ \text{M}_{\odot}$, the dashed ones represent the average over the tested parameter space. In both figures, the effective radii are expressed as comoving distances.} 
 \label{search_range}
\end{figure*}

The $R_{\text{eff}}$ calculated for the HKLV network are shown in Figure \ref{search_range}$\,$. In the most sensitive bins, centered at $m_1 = m_2 = 75 \ \mbox{M}_{\odot}$ and $m_1 = m_2 = 125 \ \mbox{M}_{\odot}$, $R_{\text{eff}} = 3 \ \mbox{Gpc}$ (redshift $z \sim 0.9$ within the cosmological model considered for this analysis, see Appendix \ref{appendix_B}). The effective radius averaged over the investigated mass values is $\langle R_{\text{eff}} \rangle \sim 1.8 \ \mbox{Gpc}$ $\left(z  \sim 0.5 \right) \,$. Hereafter, the brackets will denote the quantities averaged over the investigated $m_1$ and $m_2$ bins. The statistical error on $R_{\text{eff}}\,$, due to the finite number of injections performed, is computed as in \cite{Abadie4, Aasi3} and is $\sim 1\%\,$ over most of the tested parameter space, including the most sensitive mass bins.

The $R_{\text{eff}}$ calculated for the HLV and HL configurations agree within a few percent with those computed for the HKLV network. This is mainly due to fact that the strongest IMBHB GW emission occurs at frequencies below $\sim 10^2$ Hz. Over most of this frequency range, the advanced LIGO design sensitivity is better than advanced Virgo and KAGRA, see Figure \ref{sensitivity}$\,$. Thus, the LIGO detectors play a leading role in the sensitivity of the considered networks. Moreover, our analysis was conducted on simulated Gaussian stationary noise, and the same threshold on the reconstructed network SNR was applied to the considered networks. However, for searches on real data, higher thresholds are applied to networks including fewer detectors, leading to smaller $R_{\text{eff}}\,$. This is due to the more efficient separation of genuine GWs from noise events provided by larger networks. Thus, the reduction of the background of future searches will strongly benefit from the Virgo and KAGRA observatories. As regards the Early HL configuration, the largest $R_{\text{eff}}$ is equal to 0.7 Gpc ($z \sim 0.2$), whereas $\langle R_{\text{eff}} \rangle \sim 0.4$ Gpc ($z \sim 0.1$).

Figure \ref{search_range} shows how the largest $R_{\text{eff}}$ and the $\langle R_{\text{eff}} \rangle$ calculated for the HKLV and Early HL configurations vary over a wide range of possible thresholds on the reconstructed network SNR. The results calculated for the HLV and HL configurations agree with those shown for the HKLV network within a few percent.

\subsection{Coalescence-rate densities}
We provide an order-of-magnitude estimate of the lowest measurable IMBHB coalescence-rate density. This was calculated as
\begin{equation} \label{rate_density}
 \mathcal{R} = \frac{1}{V_{\text{vis}} \, T_{\text{obs}}} \ ,
\end{equation}
corresponding to one event observed in the analyzed observation time $T_{\text{obs}}$. For the sake of simplicity, we assumed that the IMBHB rate does not depend on redshift within $V_{\text{vis}}$. We also ignored the $(1 + z)$ factor between the time measured in the observer and source frames. As our analysis is sensitive to IMBHBs up to $z \lesssim 1$, such a factor would not impact significantly our order-of-magnitude estimate.

For the HKLV configuration, Eq. (\ref{rate_density}) gives
\begin{equation}
 \mathcal{R} \sim 9 \times 10^{-6} \left(\frac{1 \ \mbox{yr}}{T_{\text{obs}}} \right) \ \mbox{Mpc}^{-3} \ \mbox{Myr}^{-1} \ 
\end{equation}
in the most sensitive mass bin, and 
\begin{equation}
  \langle \mathcal{R} \rangle \sim 4 \times 10^{-5} \left(\frac{1 \ \mbox{yr}}{T_{\text{obs}}} \right) \ \mbox{Mpc}^{-3} \ \mbox{Myr}^{-1}
\end{equation}
when averaged over the tested parameter space. The $\mathcal{R}$ calculated for the HLV and HL networks are comparable to the values computed for the HKLV configuration. For the Early HL network, the $\mathcal{R}$ and $\langle \mathcal{R} \rangle$ are, roughly, two orders of magnitude larger. The results can be calculated at different thresholds from the $R_{\text{eff}}$ scaling shown in Figure \ref{search_range}$\,$.

IMBHB coalescence-rate densities are commonly expressed in events per globular cluster (GC) per Gyr. We converted our results into these units by assuming, for the sake of simplicity, a redshift independent GC density of $3 \ \mbox{GC} \ \mbox{Mpc}^{-3}$ \cite{Portegies2}. For the HKLV network, this gives
\begin{equation}
 \mathcal{R} \sim 3 \times 10^{-3} \left(\frac{1 \ \mbox{yr}}{T_{\text{obs}}} \right) \ \mbox{GC}^{-1} \ \mbox{Gyr}^{-1}
\end{equation}
and 
\begin{equation}
 \langle \mathcal{R} \rangle \sim 10^{-2} \left(\frac{1 \ \mbox{yr}}{T_{\text{obs}}} \right) \ \mbox{GC}^{-1} \ \mbox{Gyr}^{-1} \ .
\end{equation}
The result is about four orders of magnitude lower than the upper limits set on previous LIGO-Virgo data \cite{Aasi4,Abadie4,Aasi3} and is comparable to the theoretical estimates in \cite{Abadie5}. Finally, it is more than one order of magnitude lower than $0.1 \ \mbox{GC}^{-1} \ \mbox{Gyr}^{-1}$, the IMBHB coalescence-rate density corresponding to one event occurring in each GC within the lifetime of the cluster (assumed equal to $10 \ \text{Gyr}$). Thus, advanced detectors could provide relevant constraints on the IMBHB merger-rate density in the local Universe. 

\subsection{Observation rates}
We estimated the IMBHB observation rates with networks of advanced detectors. The estimate relies on the current theoretical models of the IMBHB formation rate in the local Universe. These suggest that an IMBHB might form via core collapse of a young and dense stellar cluster (single-cluster channel) \cite{Gurkan1} or via the merger of two clusters, both harboring one intermediate mass black hole (double-cluster channel) \cite{Amaro-Seoane1} \footnote{A further IMBHB formation channel, based on the evolution of very massive isolated stellar binaries, has been recently proposed \cite{Belczynski}. As the model was suggested after the completion of our analysis, its contribution to the observation rate was not considered for this paper.}. However, it should be kept in mind that to date no evidence for IMBHBs has been collected and that the observed rate might be zero.

Following \cite{Fregeau, Gair, Amaro-Seoane3}, we estimated $\mathcal{N}_{\text{sc}}$, the observation rate of IMBHBs formed via single-cluster channel, as
$$\mathcal{N}_{\text{sc}} = \frac{2 \times 10^{-3} \, g \, g_{\text{cl}}}{\ln( M_{\text{tot,max}} / M_{\text{tot,min}})} \int_{M_{\text{tot,min}}}^{M_{\text{tot,max}}} \frac{d M_{\text{tot}}}{M_{\text{tot}}^2} \int_{q_{\text{min}}}^{q_{\text{max}}} dq $$
$$\int_0^{z_{\text{max}} (M_{\text{tot}}, \, q)} \, dz \, 0.17 \frac{e^{3.4 z}}{e^{3.4 z} + 22} \frac{4 \pi (c/H_0)^3}{(1 + z)^{5/2}} \times $$
\begin{equation} \label{single_cluster_rate}
 \times \left\{ \int_0^z \frac{dz'}{[\Omega_M (1 + z')^3 + \Omega_{\Lambda}]^{1/2}} \right\}^2 \ .
\end{equation}
In the above equation, the meaning of the parameters is as follows:
\begin{enumerate}
 \item $M_{\text{tot}}$ and $q$ are the IMBHB total mass and mass ratio. The limits of integration depend on the investigated parameter space. For this study, $M_{\text{tot,min}} = 50 \ \mbox{M}_{\odot}$, $M_{\text{tot,max}} = 1050 \ \mbox{M}_{\odot}$, $q_{\text{min}} = 1/6$ and $q_{\text{max}} = 1\,$. $z_{\text{max}}(M_{\text{tot}}, q)$ is the maximum redshift at which the analysis is sensitive to an IMBHB with total mass $M_{\text{tot}}$ and mass ratio $q$, and was determined from the $R_{\text{eff}}$ calculated as outlined in Section \ref{analysis_overview}$\,$.
 \item $H_0$, $\Omega_M$ and $\Omega_{\Lambda}$ depend on the considered cosmological model and are defined in Appendix \ref{appendix_B}$\,$. $c$ is the speed of light.
 \item $g$ is the fraction of globular clusters in which one pair of intermediate mass black holes forms. The model relies on the assumption that the presence of more than two intermediate mass black holes in one cluster is unlikely.
 \item $g_{\text{cl}}$ is the fraction of star-forming mass hosted in the globular clusters of interest. Here $g_{\text{cl}}$ was assumed to be redshift independent.
\end{enumerate}

The values of $g$ and $g_{\text{cl}}$ are affected by large uncertainties. In the literature, $g$ and $g_{\text{cl}}$ are typically set to the fiducial value of $0.1$ \cite{Fregeau, Gair}. Simulation studies indicate that $g$ could be as large as 0.5 \cite{Freitag}, whereas observations suggest that $g_{\text{cl}}$ might be closer to $\sim 0.0025$ rather than 0.1 \cite{McLaughlin}. We set $g$ to the more conservative value of $0.1\,$, rather than $0.5\,$, and let $g_{\text{cl}}$ vary between $0.0025$ and $0.1\,$. As $\mathcal{N}_{\text{sc}}$ depends linearly on these parameters, our result can be easily rescaled at different $g$ and $g_{\text{cl}}$ values.

For the HKLV configuration, the numerical integration of Eq. (\ref{single_cluster_rate}) suggests that, roughly, 
\begin{equation}
 \mathcal{N}_{\text{sc}} \, \in \, \left[ 2, \, 80 \right] \, \text{yr}^{-1} \ ,
\end{equation}
where the interval depends on the considered range of $g_{\text{cl}}$ values. Because of the comparable $R_{\text{eff}}$, similar $\mathcal{N}_{\text{sc}}$ were calculated for the HLV and HL configurations. Our result is consistent with the rates estimated in \cite{Amaro-Seoane3} when the same normalization of $g_{\text{cl}}$ is assumed\footnote{The observation rate reported in \cite{Amaro-Seoane3} was estimated for the case of one advanced LIGO detector at the SNR threshold of 8$\,$. This threshold is not significantly different from the HL average single-detector SNR threshold applied in this paper. For a network of $N$ detectors sharing comparable sensitivities, the average reconstructed single-detector SNR is estimated by dividing the reconstructed network SNR by $\sqrt{N}$. At the threshold we considered, this yields an average HL single-detector threshold of $11/\sqrt{2} \sim 8\,$.}. Regarding the Early HL configuration, $\mathcal{N}_{\text{sc}}$ could vary within, roughly, $\left[0.02, \, 0.7 \right] \, \text{yr}^{-1}$.
 
The observation rate of IMBHBs formed via the double-cluster channel, $\mathcal{N}_{\text{dc}}$, is estimated in \cite{Amaro-Seoane1,Amaro-Seoane3} as
\begin{equation} \label{double_channel}
 \mathcal{N}_{\text{dc}} = g \, P_{\text{coll}} \, \mathcal{N}_{\text{sc}} \ .
\end{equation}
Here $P_{\text{coll}}$ is the probability of two clusters colliding. $P_{\text{coll}}$ is currently uncertain and could have values in the range $\left[0.1, 1 \right]$ \cite{Amaro-Seoane1}. Thus, the contribution to the total observation rate from the double-cluster channel could be significant for large cluster-collision probability.

\subsection{Impact of SNR loss on the ranges}

\begin{figure}[t]
 \begin{center}  
  \includegraphics{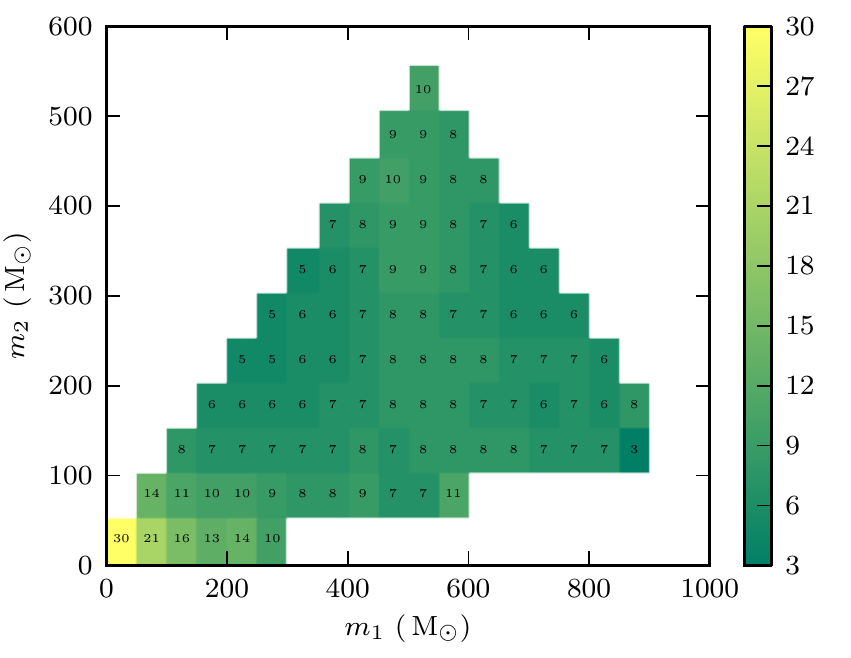}
  \caption{(color online) Percentage reduction of the coherent WaveBurst HKLV effective radii with respect to those computed by optimally matched-filtering the data at the same threshold on the network SNR used in this paper. The difference is expressed as a function of the source-frame companion masses. The coherent WaveBurst effective radii are smaller as a number of the simulated signals are not recovered due to the fraction of the injected SNR lost by the algorithm.}
  \label{effective_radius_variation}
 \end{center}  
\end{figure}

Loss of signal power could cause a number of events to be reconstructed below the applied threshold, lowering the detection efficiency. We estimated the $R_{\text{eff}}$ decrease due to the fraction of the simulated signals' SNR not recovered by coherent WaveBurst. To this aim, we calculated the coefficients
\begin{equation}
 \delta R_{\text{eff}} = \frac{R_{\text{eff, id}} - R_{\text{eff}}}{R_{\text{eff, id}}} \ ,
\end{equation}
where $R_{\text{eff, id}}$ is the range calculated by using Eq. (\ref{visible_volume}) and defining as recovered the simulated signals with injected network SNR larger than 11$\,$. The injected network SNR was calculated by summing in quadrature the injected single-detector SNR. Thus, the $R_{\text{eff, id}}$ can be interpreted as the $R_{\text{eff}}$ calculated via optimally matched-filtering the data at the network-SNR threshold adopted in this paper.

The $\delta R_{\text{eff}}$ calculated for the HKLV network vary within $\left[3\%, 10\% \right]$ over most of the tested parameter space, see Figure \ref{effective_radius_variation}$\,$. Due to the longer duration of the inspiral stage within the detectors' bandwidths, a larger difference was found in the low-mass regime $(\sim 30\%)$. Comparable results were calculated for the HLV and HL networks. We conclude that, for most of the tested binary systems, the results calculated with the unmodeled coherent WaveBurst algorithm are not significantly different from those provided, at the same SNR threshold, by an ideal reconstruction algorithm. Nevertheless, it should be kept in mind that on real data an ideal search based on optimal matched filtering would be more efficient at separating GWs from noise and would be conducted at a lower threshold. 

Further improvement is expected from the upgraded coherent WaveBurst algorithm currently under testing. Particularly relevant will be the more efficient SNR recovery over the time-frequency area spanned by the in-band portion of the signal, which should significantly increase the search sensitivity to IMBHBs in the low-mass regime. 
 \section{Sources of uncertainties on the results} \label{discussion}
Our results are affected by a number of uncertainties that are hard to estimate or model. The most relevant source of uncertainty is the sensitivity at low frequencies for the  advanced detectors. The estimates computed for the Early HL network provide a rough indication on the interval over which the results might vary depending on the eventual sensitivity. Moreover, the study was conducted on simulated data and the thresholds we applied were selected somewhat arbitrarily. The adopted threshold may be optimistic, leading to an over-estimate of the detectors' sensitivity. Nevertheless, we show in Figure \ref{search_range} how the sensitivity varies at more conservative thresholds. 

Further uncertainties arise from the spins of the IMBHB components. The amount of energy released via GWs by black-hole binaries depends strongly on the magnitude of the companion spins and on their alignment with the binary's orbital angular momentum. Compared to non-spinning binaries, a larger (smaller) amount of energy is lost to GWs by systems with aligned (anti-aligned) spins \cite{Campanelli}, increasing (decreasing) the $R_{\text{eff}}\,$. Although spinning black holes are expected to be commonplace \cite{McClintock,Reynolds}, only non-spinning companions were considered for this paper. This was due to the fact that waveforms modeling the coalescence of precessing companions, i.e., the most general spin configuration, are in progress \cite{Hannam2,Hannam:2013waveform,Pan2,Taracchini:2013,Sturani} and were not available at the time of the analysis.

Another relevant source of uncertainty is the lack of astrophysical constraints on IMBHBs. This mainly impacts the estimate of the IMBHB observation rates, which depends crucially on the adopted astrophysical models. The most critical assumptions of the model in Eq. (\ref{single_cluster_rate}) are:
 \begin{enumerate}
  \item All massive young clusters become globular clusters. However, it is currently unknown whether the initial conditions required for the formation of intermediate mass black holes in young clusters could lead to globular clusters. Moreover, the massive young clusters observed today, such as the Arches and the Quintuplet, show different properties with respect to globular clusters \cite{Figer1, Figer2}. Young clusters are less massive than globular clusters and are located in galactic disks, whereas globular clusters are harbored in the halo \cite{Portegies3}. Finally, most known massive young clusters are expected to dissolve due to the galactic tidal field in less than $\sim 1 \ \text{Gyr}$ \cite{Gieles}.
  \item The distribution of the globular-cluster total mass $M_{\text{cl}}$ scales as $(d N_{\text{cl}} / d M_{\text{cl}}) \propto M_{\text{cl}}^{-2}$ in the mass range $[10^4, 10^6] \ \mbox{M}_{\odot} \,$. Here, $dN_{\text{cl}}$ is the number of globular clusters within the mass interval $d M_{\text{cl}}$. This distribution is suggested by observations of the Antennae galaxies \cite{Zhang}. However, the Antennea galaxies are located $\sim 20$ Mpc away \cite{Schweizer}. The extension of the validity range of the globular-cluster mass distribution to the larger volumes accessible by the advanced detectors is not supported by observations and must be taken with caution. 
  \item IMBHBs are assumed to be uniformly distributed in mass ratio between 0 and 1$\,$. No observational evidence confirming this scenario is currently available.
  \item The IMBHB total mass and the mass of the host cluster scale as $2 \times 10^{-3}\,$. This relation is suggested by simulation studies \cite{Gurkan2}, but compelling observational proof is still lacking.
 \end{enumerate}
Finally, the formation rates via the single- and double-cluster channels depend on a number of parameters, such as $g$, $g_{\text{cl}}$ and $P_{\text{coll}}$. These are affected by large uncertainties and could significantly vary the estimated observation rates.
 \section{Conclusion} \label{conclusion}
We estimated the sensitivity of advanced gravitational-wave detectors to intermediate mass black hole binaries. This new class of observatories could be sensitive to the considered systems up to the Gpc scale over a wide parameter space. The measurable rate densities are expected to be a few orders of magnitude smaller than the current upper limits and sufficiently low to provide observational constraints on the theoretical models of the merger-rate density. We also found that the observation rates could be as large as few tens of events per year.

To summarize, our results suggest that advanced detectors may prove the existence of low-redshift intermediate mass black holes and open a new window for direct observations up to cosmological distances, leading to a dramatic improvement in our understanding of the Universe.
 \section*{Acknowledgments} \label{acknowledgments}
The authors thank Pau Amaro-Seoane, Thomas Dent, Ilya Mandel, Michela Mapelli, Reinhard Prix and Pablo A. Rosado for useful discussions. GM is very thankful to Bruce Allen and the International Max Planck Research School on Gravitational Wave Astronomy (IMPRS-GW). CP gratefully acknowledges support by NSF grant PHY-0970074$\,$. SK gratefully acknowledges support by NSF grant PHY-1205512$\,$. MD is thankful to EGO for the support. This document was assigned the LIGO number LIGO-P1300053$\,$.

 \appendix

 \section{Analysis thresholds} \label{appendix_A}
To better separate genuine GWs and noise events, the coherent WaveBurst triggers identified on real data are selected by thresholding three major test statistics: the coherent network amplitude ($\eta$), the network correlation coefficient ($cc$) and the network energy disbalance ($\lambda$) \cite{Klimenko1,Pankow}. The $\eta$ statistic estimates the magnitude of the event. Thus, thresholding $\eta$ has a direct impact on the search range. The $cc$ and $\lambda$ statistics assess the event consistency. The $cc$ statistic estimates the coherence of the reconstructed trigger: true GWs should be reconstructed with $cc$ close to unity, noise events with $cc \ll 1\,$. Values of $\lambda$ significantly larger than zero identify the unphysical solutions for which the reconstructed energy is larger that the energy of the data stream. 

For this analysis, only the simulated signals reconstructed with $cc > 0.7$ and $\lambda < 0.4$ were considered, consistently with the past IMBHB searches \cite{Abadie4,Aasi3}. On the contrary, a different procedure was considered to threshold the triggers' magnitude: the simulated signals were selected based on the reconstructed network SNR, rather than on $\eta$, see Section \ref{analysis_overview}$\,$. This was due to the fact that the threshold on $\eta$ is based on estimates of the search background \cite{Abadie4,Aasi3}, which were not available at the time of this analysis.

The threshold on the reconstructed network SNR used in this paper was empirically found to correspond, roughly, to the $\eta$ values in Table \ref{analysis_thresholds}$\,$. The equivalence was estimated in terms of the coefficients 
\begin{equation}
 \Delta R_{\text{eff}} = \frac{R_{\text{eff}, \eta} - R_{\text{eff}}}{R_{\text{eff}}} \ ,
\end{equation}
where $R_{\text{eff}, \eta}$ is the effective radius calculated by applying the $\eta$ thresholds in Table \ref{analysis_thresholds}$\,$. The $\Delta R_{\text{eff}}$ were found to vary, roughly, within $[-5\%, 5\%]$, depending on the network and on the considered mass bin.

\begin{table}[h]
 \begin{center}
   \begin{tabular}{cc}
      \hline 
      \hline
      \ Network \ \ \ & \ \ \ $\eta$ \ \\ 
      \hline
      \hline
      \ HKLV \ \ \ & \ \ \ 3.2 \ \\
      \ HLV \ \ \ & \ \ \ 3.7 \ \\ 
      \ HL \ \ \ & \ \ \ 4.2 \ \\ 
      \ Early HL \ \ \ & \ \ \ 4.3 \ \\
      \hline
      \hline
   \end{tabular}
 \end{center}
 \caption{Thresholds on the $\eta$ statistic empirically found to be roughly equivalent to the threshold of 11 on the reconstructed network SNR adopted for the present analysis.}
 \label{analysis_thresholds}
\end{table}
 \section{Impact of redshift on binary searches} \label{appendix_B}
For the analysis reported in this paper, a flat $\Lambda_{\text{CDM}}$ Universe was considered\footnote{New estimates of several cosmological parameters have been released after the completion of this analysis \cite{Ade}. The new values are not expected to impact significantly the results we present. The associated redshift corrections to masses and distances differ at most by few percent. This induces variations of our final results which are small compared to the uncertainties affecting our analysis and briefly discussed in Section \ref{discussion}$\,$.}, with $H_0 = 72 \ \mbox{km} \ \mbox{s}^{-1} \ \mbox{Mpc}^{-1}$ (Hubble constant), $\Omega_{M} = 0.27$ (total mass density) and $\Omega_{\Lambda} = 0.73$ (dark energy density) \cite{Komatsu}. 

Due to the expansion of the Universe, for a binary system with component masses $m_1$ and $m_2$ and located at redshift $z$, i.e., at the comoving distance $D_C$ defined as \cite{Hogg}:
\begin{equation}
 D_C = \frac{c}{H_0} \int_0^z \frac{dz'}{\sqrt{\Omega_M \left(1 + z' \right)^3 + \Omega_{\Lambda}}} \,,
\end{equation}
the corresponding quantities measured at the detector are \cite{Maggiore, Markovic}
\begin{equation}
 m_{i,\,z} = \left(1 + z \right) m_i \ \ \ \text{and} \ \ \ D_L = \left(1 + z \right) D_C \,.
\end{equation}
In the previous equations, $c$ is the speed of light, $D_L$ the luminosity distance and $i = 1,\, 2\,$. 

At the sensitivity attained by the LIGO and Virgo detectors over the past years, the redshift effects on IMBHB searches can be safely disregarded. The largest ranges were $\mathcal{O}\left(10^2 \right)$ Mpc \cite{Abbott4,Aasi4,Abadie4,Aasi3}, corresponding to $z$ values smaller than $\sim 10^{-2}$. Redshift effects were therefore smaller than other sources of uncertainties, such as the detectors' calibration \cite{Abbott4,Aasi4,Abadie4,Aasi3}. This scenario is expected to change in the advanced detector era, as the new class of observatories will be sensitive to IMBHBs up to the Gpc scale and redshift effects will, therefore, become relevant.
 \bibliography{ADE_paper}

\end{document}